\documentclass[galaxies,article,submit,pdftex,moreauthors]{mdpi} 
\firstpage{1} 
\pubvolume{1}
\issuenum{1}
\articlenumber{0}
\pubyear{2022}
\copyrightyear{2022}
\datereceived{} 
\dateaccepted{} 
\datepublished{} 
\hreflink{https://doi.org/} 

\newcommand{\msun}{M_{\odot}}

\usepackage{amsmath,amssymb}
\usepackage{mathtools}
\usepackage{bbold}
\usepackage{booktabs}
\usepackage{tensor}

\usepackage{float}

\usepackage{multirow}

\usepackage{mwe}

\usepackage{changes}

\usepackage{enumerate}

\allowdisplaybreaks

\newcommand{\vare}{\varepsilon}

\newcommand{\pt}{\partial}

\Title{Neutron star radii, deformabilities, and moments of inertia from experimental and ab initio theory constraints on the 208Pb neutron skin thickness}

\TitleCitation{Title}

\Author{Yeunhwan Lim $^{1}$ and Jeremy W. Holt$^{2}$}

\AuthorNames{Yeunhwan Lim and Jeremy W. Holt}

\AuthorCitation{Lim, Y.; Holt, J. W.}

\address{%
$^{1}$ \quad Department of Science Education, Ewha Womans University, Seoul, Korea; yeunhwan.lim@ewha.ac.kr\\
$^{2}$ \quad Cyclotron Institute and Department of Physics and Astronomy, Texas A\&M University, College Station, USA; holt@physics.tamu.edu}

\corres{Correspondence: yeunhwan.lim@ewha.ac.kr}

\abstract{Recent experimental and ab initio theory investigations of the $^{208}$Pb neutron skin thickness are sufficiently precise to inform the neutron star equation of state. In particular, the strong correlation between the $^{208}$Pb neutron skin thickness and the pressure of neutron matter at normal nuclear densities leads to modified predictions for the radii, tidal deformabilities, and moments of inertia of typical $1.4\,\msun$ neutron stars. In the present work, we study the relative impact of these recent analyses of the $^{208}$Pb neutron skin thickness on bulk properties of neutron stars within a Bayesian statistical analysis. Two models for the equation of state prior are employed in order to highlight the role of the highly uncertain high-density equation of state. From our combined Bayesian analysis of nuclear theory, nuclear experiment, and observational constraints on the dense matter equation of state, we find at the 90\% credibility level $R_{1.4}=12.36^{+0.38}_{-0.73}$\,km for the radius of a $1.4\,\msun$ neutron star, $R_{2.0}=11.96^{+0.94}_{-0.71}$\,km for the radius of a $2.0\,\msun$ neutron star, $\Lambda_{1.4}=440^{+103}_{-144}$ for the tidal deformability of a $1.4\,\msun$ neutron star, and $I_{1.338}=1.425^{+0.074}_{-0.146}\, \times 10^{45}\,\rm{g\,cm}^{2}$ for the moment of inertia of PSR J0737-3039A whose mass is $1.338\,\msun$.}


\begin{document}




\section{Introduction}

Neutron stars have attracted a great deal of attention in recent years as new measurements of masses from radio pulsar timing \cite{Demorest2010,Antoniadis2013,Cromartie2019}, tidal deformabilities from gravitational wave analyses \cite{abbott17a,abbott2018b}, and radii from x-ray pulse profiling \cite{Miller2019,Riley2019,Miller2021,Riley2021} have begun to clarify the possible existence of novel states of matter in the dense inner cores of the heaviest neutron stars and the equation of state of dense matter \cite{essick21prc,li21,Drischler:2021kxf}. At the same time, progress in nuclear effective field theories \cite{epelbaum09rmp,machleidt11} have enabled similarly strong constraints \cite{hebeler10prl,annala18,lim18a,tews2018gw,lim19epj,Capano:2019eae,Drischler:2020fvz,Drischler:2021kxf} on the dense matter equation of state and bulk properties of typical $1.4\,\msun$ neutron stars. These nuclear theory models can be further refined through precise investigations of neutron-rich nuclei, which probe the density dependence of the nuclear isospin-asymmetry energy that characterizes the energy cost of converting protons into neutrons in a nuclear many-body medium. Recently, the PREX-II experimental measurement \cite{prex2021,reed2021} of the $^{208}$Pb neutron skin thickness $R_{np} \equiv R_n - R_p = (0.283 \pm 0.071)$\,fm has provided new insights into the nuclear isospin-asymmetry energy, which can then be combined with neutron star observations in comprehensive statistical analyses \cite{essick21prc,biswas21} of the dense matter equation of state. One finds that on one side, measurements of neutron star tidal deformabilities from merging binaries and nuclear theory calculations based on chiral effective field theory tend to favor soft equations of state around $1-2$\,$n_0$, where $n_0$ is the nuclear matter saturation density. On the other side, NICER radius measurements and the PREX-II neutron skin thickness measurement of $^{208}$Pb tend to favor stiff equations of state.
Even more recently, ab initio nuclear theory calculations\,\cite{hu21} have predicted a significantly smaller neutron skin thickness $R_n-R_p = (0.14-0.20)$\,fm of $^{208}$Pb, which is consistent with the soft equations of state found from previous chiral effective field theory investigations. The goal of the present work is to systematically study the relative impact of these $^{208}$Pb neutron skin thickness analyses together with other theory, experimental, and astrophysical constraints on the dense matter equation of state, all within a Bayesian statistical framework. We will show that our posterior probability distribution on the neutron star equation of state remains relatively broad, and we find no need to introduce exotic degrees of freedom in neutron star interiors to explain present data.

One of the strongest constraints on exotic states of matter in neutron star interiors comes from neutron star mass measurements. For instance, the observation of $\sim 2\,\msun$ neutron stars \cite{Demorest2010,Antoniadis2013,Cromartie2019} strongly disfavors \cite{schulze11,Lonardoni:2014bwa} the appearance of strangeness in neutron stars at around $2-3\,n_0$, the canonical density inferred from a variety of hyperon-nucleon two-body forces \cite{rijken10,Lonardoni:2013rm}. To resolve this apparent tension, it is natural to investigate the role of the more uncertain hyperon-nucleon-nucleon three-body forces \cite{petschauer20}, which can delay \cite{Lonardoni:2014bwa,gerstung20} the onset of strangeness in neutron stars, perhaps even indefinitely. 

Observations of gravitational waves from the late-inspiral phase of GW170817 strongly favor small tidal deformabilities for $1.4\,\msun$ neutron stars, $\Lambda_{1.4} = 190^{+390}_{-120}$ \cite{abbott17a,abbott17c,abbott2018b}, and hence soft equations of state around twice nuclear matter saturation density $2n_0$. In order to infer the existence of a phase transition at high densities, it is more promising to observe gravitational waves from the post-merger phase of binary neutron star coalescence and in particular the oscillation frequencies of hypermassive neutron star remnants. Neutron star equations of state with lepton and nucleon degrees of freedom only show a strong correlation between the peak frequency of the oscillating remnant and the radius or tidal deformability of typical $1.4\,\msun$ neutron stars \cite{bauswein19}. Strong deviations from this speculated correlation could indicate the presence of a phase transition to a hybrid quark-hadron star. For most neutron star merger events, however, current gravitational wave detectors do not have sufficient sensitivity in the high-frequency band above 1kHz to measure remnant oscillations.

Neutron star radii and tidal deformabilities are strongly correlated. From GW170817 one can already place an upper bound $R_{1.4} \lesssim 13.5$\,km on the radius of $1.4\,\msun$ neutron stars \cite{annala18,abbott2018b,de2018a}. More direct measurements of neutron star radii from x-ray pulse profiling \cite{Riley2019,Miller2019,Riley2021,Miller2021} are consistent with inferences from GW170817, but statistically favor somewhat larger radii. In the case of the NICER x-ray measurements of PSR J0030+0451, two independent analyses found $M = 1.44^{+0.15}_{-0.14}\msun$ and $R = 13.02^{+1.24}_{-1.06}$\,km \cite{Miller2019} and $M = 1.34^{+0.15}_{-0.16}\msun$ and $R = 12.71^{+1.14}_{-1.19}$\,km \cite{Riley2019} at the 68\% credibility level. Furthermore, the recent results from NICER-XMM for PSR J0740+6620 also support the stiff EOS scenario. The radius of this $2.08\,\msun$ neutron star is predicted to be $R_{2.08}=13.7^{+2.6}_{-1.5}\,\rm{km}$ from Miller {\it et al.}\cite{Miller2021} and $R_{2.08}=12.39^{+1.30}_{-0.98}\,\rm{km}$ from Riely {\it et al.}\,\cite{Riley2021}. One possibility to infer the existence of a phase transition at high density from radius measurements is to search for so called ``twin stars'', that is, two neutron stars with similar gravitational mass but different radii, which would provide strong evidence for a third stable branch of cold dense stars (after white dwarfs and neutron stars) \cite{itoh70}. Present uncertainties in neutron star radius measurements, however, are likely still too large to be able to resolve hypothetical twin stars.

Complementary to the progress that has taken place in neutron star observations and astrophysics simulations, theoretical modeling of neutron stars based on state-of-the-art nuclear forces have reached sufficient accuracy to strongly constrain the neutron star equation of state and the associated properties of light $\sim 1.4\,\msun$ neutron stars \cite{lim18a,tews2018gw}. For instance, in Ref.\ \cite{hebeler10prl} chiral effective field theory calculations of the neutron matter equation of state up to nuclear matter saturation density $n_0$ were used to predict that the radius of a $1.4\,\msun$ neutron star lies in the range $9.7\,{\rm km}<R_{1.4}<13.9\,{\rm km}$. Later, in Ref.\ \cite{lim18a} it was shown that by combining chiral effective field theory priors on the dense matter equation of state together with likelihood functions obtained from isovector properties of medium-mass and heavy nuclei, the tidal deformability for $\sim 1.4\,\msun$ neutron stars lies in the range $140 < \Lambda_{1.4} < 520$ (see also Ref.\ \cite{tews2018gw} for a similar study of the binary tidal deformability parameter $\tilde \Lambda$ associated with GW170817). Low-density constraints from chiral effective field theory have also enabled more refined predictions of neutron star radii \cite{tews18kmu,lim19epj} and moments of inertia \cite{lim19prc,greif20}.

Isovector properties of medium-mass and heavy nuclei provide a third avenue to infer the empirical parameters of the nuclear equation of state around the saturation density $n_0$. For instance, the combination of experimentally measured masses, neutron skins, giant dipole resonances, and dipole polarizabilities has provided a strong constraint \cite{tsang12,lattimer13} on the nuclear isospin-asymmetry energy, with uncertainties comparable to that of recent chiral effective field theory calculations \cite{holt17prc,somasundaram20,sammarruca21}. In addition, heavy-ion collisions \cite{danielewicz02,shetty04,estee21} can give insights into the equation of state at supra-saturation densities. Recently, the measurement \cite{prex2021} of the parity-violating asymmetry in the elastic scattering of longitudinally polarized electrons off $^{208}$Pb has provided new constraints \cite{reed2021,reinhard21} on the neutron skin thickness of $^{208}$Pb and the symmetry energy slope parameter $L$, though the extraction of the latter quantity requires model assumptions. For instance, Refs.\ \cite{prex2021,reed2021} report the constraints $R_n-R_p = 0.283 \pm 0.071$\,fm and $L = 106 \pm 37$\,MeV, while Ref.\ \cite{reinhard21} reports $R_n-R_p = 0.19 \pm 0.02$\,fm and $L = 54 \pm 8$\,MeV using a wider class of energy density functionals and including also the experimentally measured electric dipole polarizability of $^{208}$Pb as a constraint. In addition, the recent ab initio nuclear theory calculation \cite{hu21} of the neutron skin thickness of $^{208}$Pb has obtained $R_n-R_p = 0.14-0.20$\,fm, which is much closer to the experimental evaluation in Ref.\ \cite{reinhard21}.

In light of these recent novel nuclear experiments, ab initio nuclear theory calculations, and neutron star observations, we study their relative impact on the dense matter equation of state and derived properties of neutron stars. Given the highly uncertain nature of the composition and pressure of ultra-dense matter beyond about twice nuclear saturation density, we employ two different prior probability distributions in our analysis. Both are based the work of Refs.\ \cite{lim18a,lim21prcl}, where chiral EFT calculations at varying cutoff scale and order in the chiral expansion are first used to construct a probability distribution of energy density functionals. These models are then further constrained by empirical information on nuclear matter properties around saturation density inferred from studies of medium-mass and heavy nuclei. Up to twice nuclear matter saturation density, our two prior distributions are identical. Beyond $2n_0$, our first prior probability distribution assumes a smooth extrapolation of the functional form obtained at lower densities. Our second prior model assumes a transition to the maximally-stiff equation of state, where the speed of sound is equal to the speed of light, at a transition density $n_c$ that is uniformly varied in the range $2n_0<n_c<4n_0$. Taken together, the two models may be used to infer potential breakdowns in the chiral effective field theory description of nuclear interactions and the possible existence of phase transitions at supra-saturation density.

The paper is organized as follows. In Section \ref{nsm} we provide a description of our neutron star modeling, which explicitly accounts for the physics of the outer and inner crusts. In Section \ref{stat} we describe details of the Bayesian statistical modeling used to account for constraints from ab initio nuclear theory, nuclear experiment, and neutron star observations. In Section \ref{sec:res} we present results for the nuclear equation of state and derived neutron star properties based on our propagated likelihood functions. We end with a summary and discussion.

\section{Materials and Methods}
\label{nsm}

\subsection{Neutron star modeling}

Many qualitative features of neutron star structure are well understood and do not depend sensitively on the choice of nuclear force model used to generate the equation of state. The neutron star crust consists of ionized heavy nuclei with the possibility of unbound neutrons beyond a transition density set by the neutron chemical potential. This inhomogeneous phase of nuclear matter exists until the total baryon number density increases to approximately one-half saturation density \cite{lim17}, at which point the lattice structure melts to a fluid of protons, neutrons, and electrons, a state of matter that has a lower energy than the inhomogeneous phase. The effects of three-nucleon forces already begin to play a role close to the crust-core transition density, and the uncertainties in chiral EFT calculations of uniform neutron matter begin to increase considerably at around half saturation density, regardless of the many-body method employed (e.g., many-body perturbation theory\,\cite{hebeler10,drischler16,drischler16b,wellenhofer16,holt17prc,drischler19},
quantum Monte Carlo\,\cite{gezerlis13,roggero14,wlazlowski14,tews16}, or self-consistent Green's function theory\,\cite{carbone14}). In addition, uncertainties in the symmetric nuclear matter equation of state, especially regarding the saturation energy and density, impact the crust-core transition density. However, the transition density does not depend on the mass of the neutron star.

The region of a neutron star below the crust is traditionally divided into an ``outer core'' and ``inner core'', where the transition to the inner core is accompanied by the appearance of novel degrees of freedom beyond protons, neutrons, electrons, and muons (the latter arise when the electron chemical potential surpasses the rest mass of the muon, $105.7$\,MeV). For instance, in the inner core, the ground state of dense matter might be comprised of hyperons, meson condensates, or deconfined quarks. Since the existence of such exotic states of matter is not certain, especially in light of observed heavy $\sim 2\,\msun$ neutron stars, the neutron star inner core remains highly speculative.

Although the general composition of the neutron star outer core is well understood, its detailed properties (e.g., proton fraction or pressure as a function of density) depend sensitively on the nuclear force. In the outer core, the nuclear equation of state can be constrained from nuclear experiments, nuclear theory, and astrophysical observations. Currently, there are more than 2300 nuclei whose masses are measured\,\cite{ame2021} and which provide a strong constraint on the isospin-asymmetry energy of the nuclear equation of state. Thus, the binding energy of finite nuclei is an essential reference together with the symmetric nuclear matter properties, the saturation energy $B$, saturation density $n_0$, and incompressibility $K$. Pure neutron matter cannot be studied directly in the laboratory, but at low to moderate densities up to $\sim 2n_0$ chiral effective field theory can provide strong constraints on the neutron matter equation of state \cite{hebeler10,coraggio13,holt17prc}. In the following we will derive properties of neutron stars through parametric modeling of the equation of state from chiral effective field theory and nuclear experiments.

\subsubsection{Neutron star equation of state}

The mass-radius relation for non-rotating neutron stars can be obtained by solving the TOV equations,
\begin{equation}
\begin{aligned}
\frac{dp}{dr} & = - \frac{[M(r) + 4\pi r^3 p](\varepsilon + p)}{r[r-2M(r)]} \,,\\
\frac{dM}{dr} & = 4\pi \varepsilon r^2\,,
\end{aligned}
\end{equation}
where $p$ and $\varepsilon$ are the pressure and total energy density,
$r$ is the distance from the center, and
$M(r)$ is the enclosed mass of a neutron star at the distance $r$ from the center. Note that we use units where $G=1$ and $c=1$. The nuclear physics inputs are therefore the energy density $\varepsilon$ and pressure $p$, which are related through the equation of state. Since the pressure is given by the derivative of the energy density with respect to the baryon number density, we can regard the EOS as the energy density for a given baryon number density. 

In principle, one can approximate the equation of state of isospin-symmetric nuclear matter through a Taylor series expansion around saturation density $n_0$:
\begin{equation}\label{eq:a0expansion}
	A_0(n) = - B + \frac{K}{9}\left(\frac{n-n_0}{n_0} \right)^2 + \frac{Q}{27}\left(\frac{n-n_0}{n_0} \right)^3 + \cdots,
\end{equation}
where $B$ is the (positive) binding energy of nuclear matter at the saturation density, $K$ is the incompressibility, and $Q$ is the skewness parameter. One may similarly expand the equation of state for asymmetric nuclear matter about the isospin-symmetric point, however, in addition to the normal even powers of the isospin asymmetry $\delta_{np} = (n_n-n_p) / (n_n + n_p)$ one also finds \cite{kaiser15,wellenhofer16} nonanalytic logarithmic terms of the form
\begin{equation}\label{eq:expansion}
	\begin{aligned}
	\frac{E}{N}(n,\delta_{np}) & = A_0(n)+S_2(n)\delta_{np}^2 \\
& 
+ \sum_{i=2}^{\infty}(S_{2i}(n) + L_{2i}(n)\ln\vert \delta_{np}\vert)\delta_{np}^{2n}
	\end{aligned}
\end{equation}
when the equation of state is computed at least to second order in perturbation theory. Although the above expansion in Eq.\ \eqref{eq:expansion} was originally derived \cite{kaiser15} assuming a zero-range contact interaction, detailed calculations \cite{wen2021} have demonstrated that realistic nuclear forces do not give rise to any additional nonanalytic terms. One also finds \cite{wellenhofer16,lagaris81,Bombaci91,drischler16b,wen2021} that the expansion in Eq.\ \eqref{eq:expansion} is dominated by the leading term $S_2(n)$, referred to as the isospin-asymmetry parameter, which like Eq.\ \eqref{eq:a0expansion} can be expanded about nuclear saturation density:
\begin{equation}\label{eq:s2expansion}
	\begin{aligned}
			S_2(n) & = J +  \frac{L}{3}\left(\frac{n-n_0}{n_0}\right) 
		+ \frac{K_{\rm sym}}{9}\left(\frac{n-n_0}{n_0}\right)^2 \\
		 & + \frac{Q_{\rm sym}}{27}\left(\frac{n-n_0}{n_0}\right)^3
		+ \cdots\,.
	\end{aligned}
\end{equation}

Although useful for orientation, neither Eq.\ \eqref{eq:a0expansion} nor \eqref{eq:s2expansion} is generally well converged at high densities beyond $2n_0$. Thus, we model the equation of state of homogeneous matter using energy density functionals, where the energy density is given as a function of the baryon number density $n$ and proton fraction $x$:
 \begin{equation}
 	\begin{aligned}
 		 	\mathcal{E}(n,x) & = \frac{1}{2m}\tau_n + \frac{1}{2m}\tau_p \\
 		 	& + (1-2x)^2 f_n(n)  + [1-(1-2x)^2]f_s(n),
 	\end{aligned}
 \end{equation}
where
\begin{equation}
\label{eq:abparam}
	f_n(n) = \sum_{i=0}^3 a_i n^{2+i/3}\,,
	\quad
	f_s(n) = \sum_{i=0}^3 b_i n^{2+i/3}\,.
\end{equation}
In the above functional, $f_n$ ($f_s$) represents the potential energy for pure neutron matter (symmetric nuclear matter). Note that $n^{1/3} \sim k_F$ is used for the expansion instead of integer powers of $n$ 
for the smooth fit to microscopic calculations.

In chiral effective field theory, the short-distance part of the nuclear force is encoded in a set of two-nucleon (2N) and three-nucleon (3N) low-energy constants that are typically fitted to 2N and 3N scattering, reaction, and bound-state observables only. Therefore, theoretical predictions for the dense matter equation of state from chiral EFT can be used to generate a prior probability distribution for the parameters $\{a_i,b_i\}$ entering in Eq.\ \eqref{eq:abparam}. In Ref.\,\cite{lim18a} we have used a set of five chiral interactions with varying momentum-space cutoff and order in the chiral expansion to calculate the neutron matter and symmetric nuclear matter equations of state at varying orders in many-body perturbation theory (up to third order \cite{holt17prc}). The functional form in Eq.\ \eqref{eq:abparam} including four terms ($i=0,1,2,3$) in the power series was shown to be sufficient to fit the nuclear equation of state up to $2n_0$. Only small variations in the fitted parameters were observed when the maximal density was reduced to $1.5n_0$.

The binding energy of symmetric nuclear matter around the saturation density is a fine-tuned quantity, and therefore chiral effective field theory calculations tend to have sizeable uncertainties at and above saturation density. In particular, the tensor force generated by one-pion-exchange is a dominant feature in the spin-triplet interaction that contributes strongly to the uncertainty in the symmetric nuclear matter equation of state. In contrast, the pure neutron matter equation of state receives only small contributions from the tensor force, since the $L=0$ spin-triplet state is Pauli forbidden. One finds that the uncertainties in the symmetric nuclear matter and pure neutron matter equations of state around the saturation density are comparable, despite the fact that one may naively anticipate a poorer convergence of the neutron matter calculations since the neutron Fermi momentum is about 25\% larger than in symmetric nuclear matter.

The probability distribution for the equation of state parameters in Eq.\ \eqref{eq:abparam} can be significantly reduced by combining the prior distribution with likelihood functions that include experimental data from medium-mass and heavy nuclei. For this purpose, we take a set of 205 mean field models \cite{dutra12} fitted to the properties of finite nuclei and compute the derived constraints on the nuclear matter empirical parameters. In the case of the prior probability distribution from chiral effective field theory, we do not consider correlations between the sets of $\{a_i\}$ and $\{b_i\}$ parameters, since our aim is to generate a relatively conservative estimate of the distribution. In contrast, recent work \cite{drischler20l} has shown that such correlations can have an important impact on uncertainty estimates for derived quantities, such as the nuclear isospin-asymmetry energy.

\subsubsection{Uniform nuclear matter}

Our posterior probability distribution for the equation of state parameterization in Eq.\ \eqref{eq:abparam} is only constrained in a relatively narrow region $0<n<2n_0$, where both chiral effective field theory calculations and finite nuclei properties are informative. However, the central density in the heaviest neutron stars can reach $5-10n_0$, and therefore we consider two separate models for the high-density equation of state. First, our ``smooth extrapolation'' extends the posterior probability distribution for the equation of state to the highest densities without modification. The only additional constraint we impose on this distribution is that the speed of sound cannot exceed the speed of light (such unphysical equations of state are removed from our modeling). Second, our ``maximally stiff extrapolation'' \cite{lim21prcl} probes the limiting case where the speed of sound achieves the maximal value of $c_s=c$ at and above a specified transition density $n_c$. This maximally stiff extrapolation allows one to probe the maximum neutron star mass consistent with low- to moderate-density constraints from nuclear physics and to draw model-independent conclusions about the density at which traditional nuclear physics models break down.

For the maximally stiff high-density extrapolation, we introduce a smooth jump in the speed of sound at the critical density $n_c$,
\begin{equation}\label{eq:cs2}
	\frac{c_s^2}{c^2} = 
	\begin{cases}
		1 & \text{if} \quad  \varepsilon > \varepsilon_c + \Delta_\varepsilon \,, \\
		\tilde{c}_c^2 + \frac{1-\tilde{c}_c^2}{\Delta_\vare}(\vare-\varepsilon_c)  
		   & \text{if} \quad  \varepsilon_c < \varepsilon < \varepsilon_c + \Delta_\varepsilon \,, \\
		\chi\text{EFT}   & \text{otherwise}\,,
	\end{cases}
\end{equation}
where $\varepsilon_c $ is the energy density at $n=n_c$ and $\tilde{c}_c^2$ is the speed of sound at $n=n_c$ in units of $c^2$. The parameter $\Delta_\vare$ is set to $\frac{1}{10}\vare_c$ for a rather smooth transition. In this work, we vary $n_c$ from $2n_0$ to $4n_0$ to simulate the effects on the neutron star mass-radius relation, tidal deformabilities, and moments of inertia. From the definition of the speed of sound $c_s^2/c^2 = \frac{\pt p}{\pt \vare}$, we can obtain the pressure and baryon number density, which is necessary for solving the TOV equations. The pressure in the phase transition region, i.e., between $\vare_c$ and $\vare_c + \Delta_\vare$, is given by
\begin{equation}
	p = p_c + \tilde{c}_c^2(\vare -\vare_c) + \frac{1-\tilde{c}_c^2}{2\Delta_\vare}(\vare-\vare_c)^2\,.
\end{equation}
Note that $\vare$ is an independent variable, and the pressure is simply determined by the relation between $\tilde{c}_c^2$ and $\vare$.

\subsubsection{Neutron star crust}
The neutron star crust consists of nuclear clusters, i.e., ionized heavy nuclei, which have a lower ground state energy than homogeneous matter at the same density. In addition, neutrons may drip out of heavy nuclei as the total baryon number density increases or more nucleons are added to a fixed volume of box. In this case, the total energy density ($F_{\rm tot}$) consists of the internal energy of heavy nuclei ($F_H$), the internal energy of neutrons ($F_o$), the electron contribution ($F_e$), the Coulomb interaction ($F_C$) between protons and electrons, and the surface energy ($F_{\rm Surf}$) to make the non-uniform nuclear density profile. Therefore, the ground state configuration is obtained by minimizing the total energy density for a given baryon number density;
\begin{equation}
	\begin{aligned}
	F_{\rm{tot}} & = F_H + F_C + F_{\rm{Surf}} + F_{o} + F_{e} \\
                       & + \lambda_1[n - u n_i -(1-u)n_{no}] \\
                       & + \lambda_2[nY_p - u n_i x_i] + \lambda_3[nY_p - n_e]\, ,
	\end{aligned}
\end{equation}
where $u$ is the volume fraction of a heavy nucleus in the Wigner-Seitz cell for the numerical calculation, $n$ ($n_e$) is the total baryon (electron) number density in the Wigner-Seitz cell, $n_i$ is the baryon number density of a heavy nucleus, $Y_p$ is the proton fraction in the Wigner-Seitz cell, $n_{no}$ is the neutron number density, and $\lambda_1$, $\lambda_2$, and $\lambda_3$ are the Lagrange multipliers necessary for the energy minimization \cite{lim17}. In detail, the contributions to the free energy density are given as
\begin{equation}
\begin{aligned}
    F_{H} & = u n_i f_i,     \\
    F_{C} & = 2\pi (n_i x_i e r_N)^2 u f_d(u), \\
    F_{\rm Surf} & =\frac{\sigma(x_i)ud}{r_N}, \\
    F_o & = (1-u)n_{no}f_o, \\
    F_e & = \frac{m_e^4 c^5}{8\pi^2}\chi(x_e)\,. \\
\end{aligned}
\end{equation}
In the electron energy density, $x_e= \frac{p_{f_e}}{m_e c}$, 
$p_{f_e}$ is the electron Fermi momentum and $m_e$ is the mass of electron. The function
$\chi(x_e)$ is given by
\begin{equation}
        \chi(x_e) = \left\{
    x_e(1+x_2^2)^{1/2}(1+2x^2) - \ln [x+(1+x^2)^{1/2}]
    \right\}.
\end{equation}
At equilibrium all quantities are obtained by solving $\frac{\pt F_{\rm tot}}{\pt u}=0$,
$\frac{\pt F_{\rm tot}}{\pt n_i}=0$,
$\frac{\pt F_{\rm tot}}{\pt x_i}=0$
$\frac{\pt F_{\rm tot}}{\pt Y_p}=0$,
$\frac{\pt F_{\rm tot}}{\pt n_{no}}=0$,
$\frac{\pt F_{\rm tot}}{\pt n_{e}}=0$,
$\frac{\pt F_{\rm tot}}{\pt \lambda_1}=0$,
$\frac{\pt F_{\rm tot}}{\pt \lambda_2}=0$, and
$\frac{\pt F_{\rm tot}}{\pt \lambda_3}=0$.
Note that the energy density involving the Coloumb and surface energy can be combined into one energy density $F_{CS} = F_C +F_{\rm 
Surf}$ from nuclear virial theorem ($2F_C = F_{\rm Surf}$)\,\cite{lseos}.
The transition from inhomogeneous matter to uniform nuclear matter is found by comparing the energy density differences in this formalism.

\subsection{Bayesian statistical analysis}
\label{stat}
Neutron star properties such as radii, tidal deformabilities, and moments of inertia can be constrained by nuclear theory calculations, nuclear experiments, and astrophysical observations. Bayes' theorem states \begin{equation}
    P(M_i\vert D) = \frac{P(D\vert M_i)P(M_i)}
    {\sum_{j}P(D\vert M_j)P(M_j)}
\end{equation}
where $M_i$ are the equation of state model parameters, i.e., the set of $\{a_i\}$ and $\{b_i\}$ in Eq.\ \eqref{eq:abparam}, $P(M_i)$ is the prior probability distribution for $\{a_i\}$ and $\{b_i\}$, $D$ is the data which is independent of $\{a_i\}$ and $\{b_i\}$, and $P(D\vert M_i)$ is the likelihood. Here the independence implies that we assume no correlation between $D$ and $M_i$ before we obtain the final posterior distribution. In the present work, we implement the following data $D = \{$PREX-II, $R_{np}^{\rm th}(^{208}{\rm Pb})$, $\rm{GW170817}$, NICER I, NICER II\}, which correspond respectively to the neutron skin thickness measurement from Ref.\ \cite{prex2021}, the ab initio neutron skin thickness prediction from Ref.\ \cite{hu21}, the neutron star tidal deformability constraints from GW170817 \cite{abbott2018b}, the simultaneous mass-radius measurement of PSR J0030+0451 \cite{Riley2019,Miller2019}, and the NICER-XMM radius measurement of PSR J0740+6620 \cite{Riley2021,Miller2021}.

For the $R_{np}^{\rm exp}(^{208}{\rm Pb})$ PREX-II measurement, we use the constraint on the slope parameter of the nuclear isospin-asymmetry energy from Reed {\it et al.}\,\cite{reed2021}:
\begin{equation}
    \mathcal{L}^{\rm PREX-II} = \frac{1}{\sqrt{2\pi} \sigma_L}
    \exp\left[ -\frac{(L-\langle L \rangle)^2}{2\sigma_L^2} \right]\,,
\end{equation}
where $\langle L \rangle=106$\,MeV and $\sigma_L=37$\,MeV. Although the central value of $L$ from the analysis in Ref.\ \cite{reed2021} appears inconsistent with previous analyses \cite{lattimer13,tews17,lim19epj,gandolfi2012,drischler20l,drischler20l,essick21prc}, the $R_{np}^{\rm exp}(^{208}{\rm Pb})$ measurement and model-dependent extraction of $L$ show large uncertainties, i.e., $\sigma_L=37$\,MeV. Thus the $95\%$ confidence interval ($32\,\rm{MeV} < L < 180\,\rm{MeV}$) covers almost all previous results. In comparison, the ab initio theory prediction $R_{np}^{\rm th}(^{208}{\rm Pb})$ of the neutron skin thickness is linked with various nuclear force parameters so that nuclear matter properties such as the binding energy, saturation density, incompressibility, symmetry energy, and its slope parameters are correlated. In their results, the symmetry energy is defined as the energy difference between pure neutron matter and symmetric nuclear matter at the nuclear saturation density, which in their calculations can vary between $0.14-0.18$\,fm$^{-3}$. Thus it is necessary to construct a 3-dimensional kernal density estimate (KDE) for a given set of density, symmetry energy, and slope parameter with corresponding likelihood
\begin{equation}
    \mathcal{L}^{R_{np}^{\rm th}(^{208}\rm Pb)} = P(n_0, S_v, L)\,,
\end{equation}
where $P(n_0, S_v, L)$ is the 3D-KDE from the analysis of Hu {\it et al.}\,\cite{hu21}.

\begin{figure*}[t!]
\begin{adjustwidth}{-\extralength}{0cm}
    \centering
    \includegraphics[width=18.5cm]{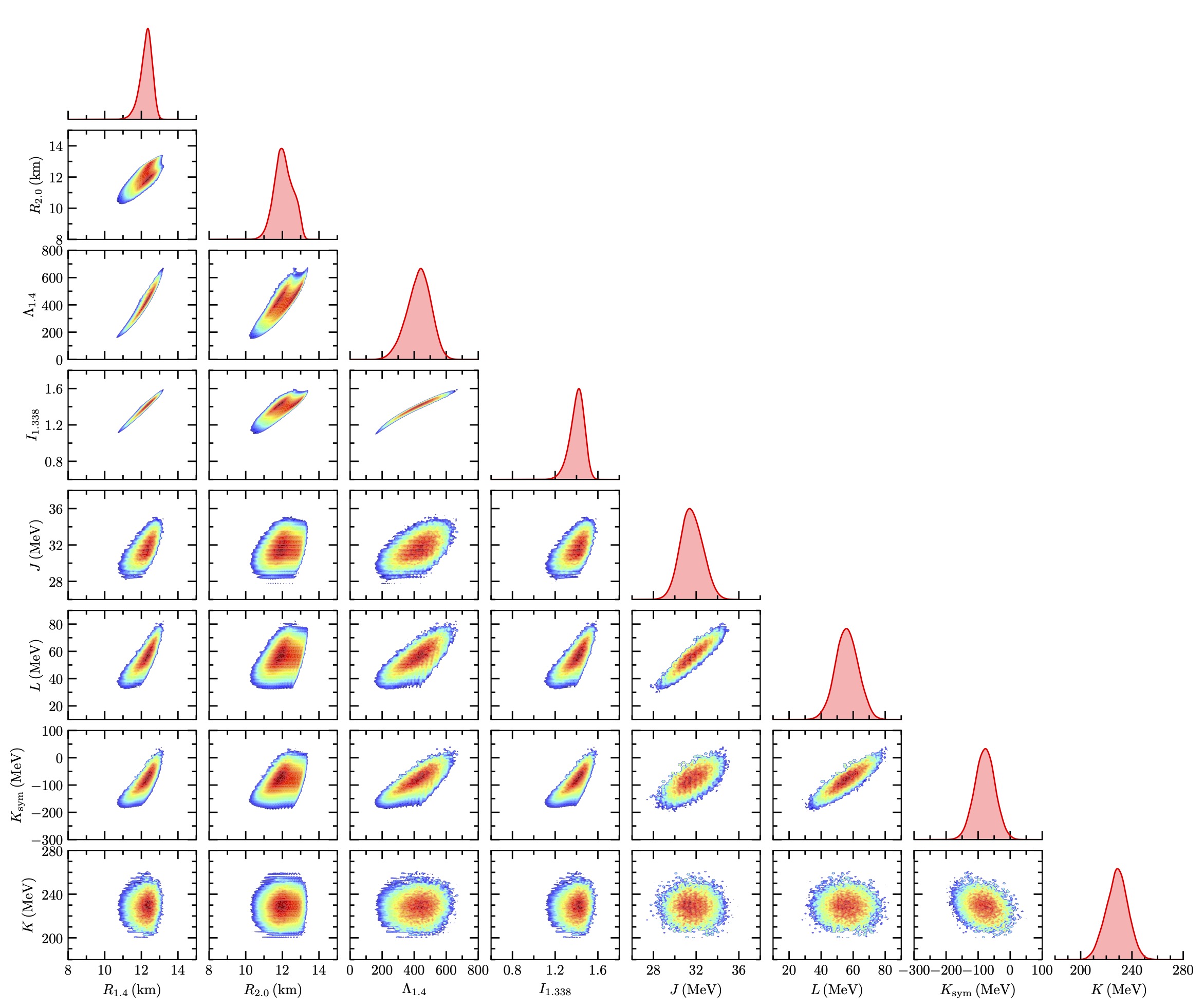}
    \caption{Correlations among neutron star properties and nuclear matter properties from the posterior probability distribution Eq.\ \eqref{eq:finpost} when the maximally-stiff high-density EOS extrapolation is used for $\mathcal{L}^{\rm prior}$.}
    \label{fig:corrori}
\end{adjustwidth}
\end{figure*}

On the astrophysical side, analysis of the gravitational waves from the neutron star merger event GW170817\,\cite{ligo19prx} provides a likelihood by integrating the allowed range of the neutron star mass for a given neutron star EOS,
\begin{equation}
    \mathcal{L}^{GW} = \int f(M_1,\Lambda_1,M_2,\Lambda_2)\, dM_1dM_2\,,
\end{equation}
where $f(M_1,\Lambda_1,M_2,\Lambda_2)$ is a four-dimensional probability. 
Since the EDF employed in this work determines the tidal deformability $\Lambda$,
it is not necessary to perform the $\Lambda_1$ and $\Lambda_2$ integrals. The results from NICER I \cite{Miller2019,Riley2019} and NICER II \cite{Miller2021,Riley2021} provide posterior samples from the joint mass-radius distribution $f(M,R)$ that enters in the likelihood:
\begin{equation}
    \mathcal{L}^{\rm NICER I, II} = \frac{1}{M_{\rm max} - M_{\rm start}}
    \int_{M_{\rm start}}^{M_{\rm max}} f(M,R) \,dM\, ,
\end{equation}
where $M_{\rm max}$ stands for the maximum mass of a neutron star for a given EOS. We set  the starting mass $M_{\rm start}$ of a neutron star for the integration as $1M_{\odot}$, since the mass distribution of observed neutron stars has no weight below this value. In principle, another likelihood could come from observed heavy neutron stars \cite{Demorest2010,Antoniadis2013}. However, the NICER analysis of PSR J0740+6620 \cite{Miller2021, Riley2021} already places a strong constraint on the maximum neutron star mass. The final posterior probability is then given by
\begin{equation}
    \mathcal{L} = 
    \mathcal{L}^{\rm PREX-II}
    \mathcal{L}^{R_{np}^{\rm th}(^{208}\rm Pb)}
    \mathcal{L}^{\rm GW}
    \mathcal{L}^{\rm NICER I}
    \mathcal{L}^{\rm NICER II}\mathcal{L}^{\rm prior}\,.
    \label{eq:finpost}
\end{equation}
From the above equation, we can obtain the probability or weight factor for each $(\vec{a},\vec{b})$ associated with the neutron star equation of state.

\begin{figure*}[t!]
\begin{adjustwidth}{-\extralength}{0cm}
    \centering
    \includegraphics[width=8.75cm]{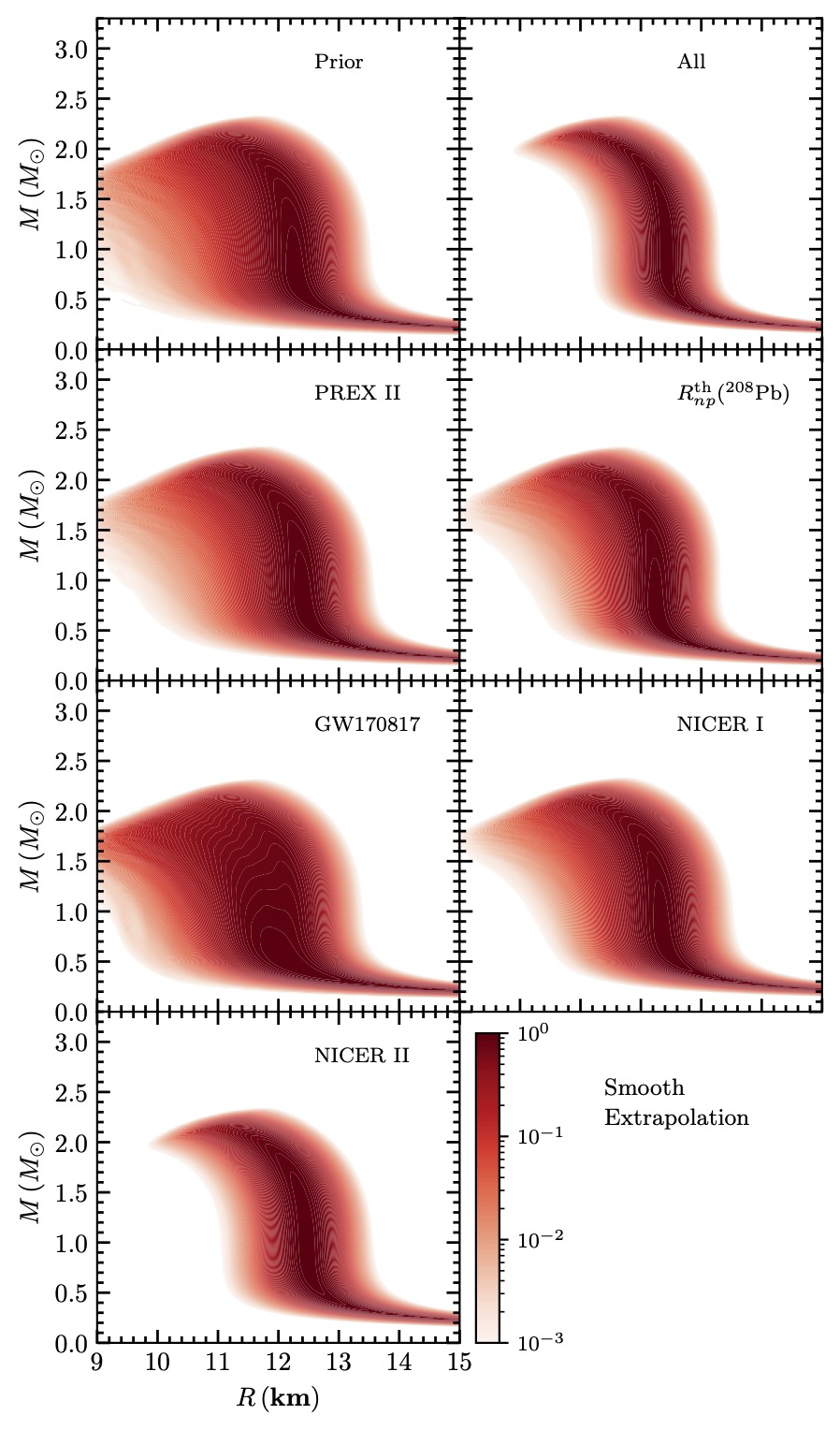}
    \includegraphics[width=8.75cm]{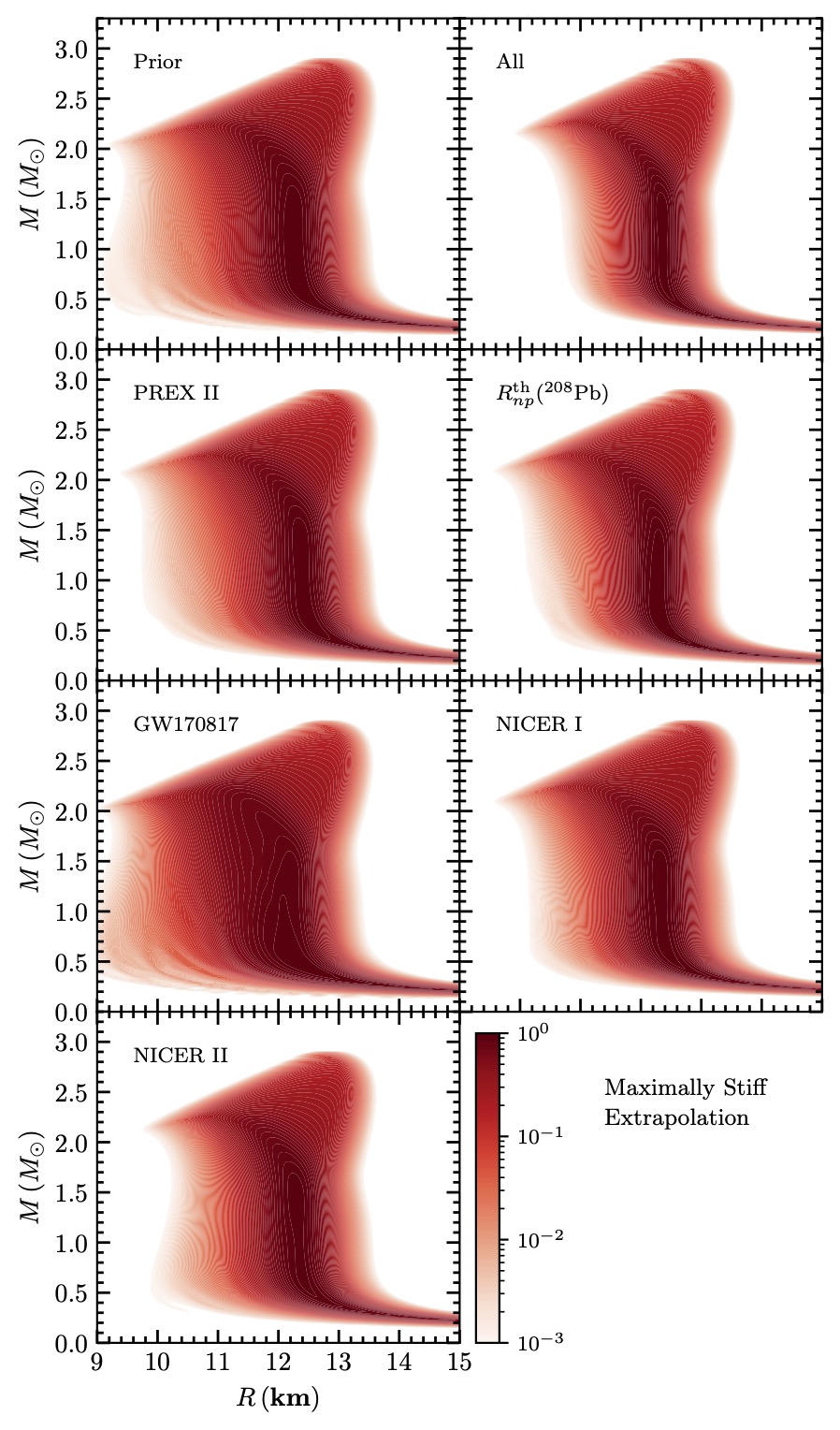}
    \caption{The mass-radius probability densities (priors, individual posteriors, and combined posteriors) from (a) the smooth high-density extrapolation and (b) the maximally-stiff high-density extrapolation energy density functionals.}
    \label{fig:mrall}
\end{adjustwidth}
\end{figure*}

\section{Results}
\label{sec:res}

The posterior distribution of EDF parameters enables us to obtain various neutron star properties as well as nuclear matter properties. Figure \ref{fig:corrori} shows the correlations among the nuclear matter properties and neutron star properties from the posterior distribution in Eq.\ \eqref{eq:finpost}, assuming the maximally-stiff high-density extrapolation for $\mathcal{L}^{\rm prior}$. At the top of each column in Figure \ref{fig:corrori} is the marginal probability distribution function associated with each of the given quantities. For example, the probability distribution for $R_{1.4}$ has a peak around 12.4\,km and the probability decreases rapidly as $R_{1.4}$ increases. The left tail, on the other hand, reaches to 11.63\,km which is the left bound for the 90\% credibility. Other plots in the first column show the correlation between $R_{1.4}$ and other neutron star properties and nuclear matter properties. In particular, one can see the very strong correlation between $\Lambda_{1.4}$ and $R_{1.4}$ which has already been analyzed in Refs.\ \cite{annala18,lim18a}. An even stronger correlation can be found between $R_{1.4}$ and $I_{1.338}$. Since the central densities for a $1.4\,\msun$ and a $1.338\,\msun$ neutron star are very close \cite{lim19epj}, the small difference in mass does not strongly affect the correlation. Both $L$ and $K_{sym}$ have strong correlations with $R_{1.4}$ and to a lesser extent $\Lambda_{1.4}$ and $I_{1.338}$. In contrast, the nuclear symmetry energy $J$ is only weakly correlated with the bulk parameters of $\sim 1.4\,\msun$ neutron stars. We find that the radius of $\sim2\,\msun$ neutron stars is not strongly correlated with any nuclear matter empirical parameter but instead can be constrained from more precise observations of radii, tidal deformabilities, or moments of inertia of $\sim 1.4\,\msun$ neutron stars.

In Figure\,\ref{fig:mrall} we show the mass-radius probability distributions under the smooth high-density extrapolation (left subfigure) and the maximally-stiff high-density extrapolation (right subfigure). The subpanels within each subfigure show the constraints imposed on the mass-radius relation from each of the separate likelihood functions in Eq.\ \eqref{eq:finpost} as well as the combined posterior distribution labeled ``All''. We see that when all likelihood functions are implemented, the mass-radius confidence interval is highly reduced. In particular, the NICER II constraint (which simultaneously enforces a lower bound on the maximum neutron star mass) has a strong impact on the posterior mass-radius relation in the ``smooth extrapolation'' scenario. This is due to the fact that chiral effective field theory predicts relatively soft equations of state, many of which lead to maximum neutron star masses less than $2\,\msun$. For the ``maximally stiff" extrapolation, the NICER II constraint also eliminates the softest equations of state, but the stiff extrapolation naturally leads to heavier neutron stars more consistent with NICER II. The strongest reductions in the space of stiff equations of state come from GW170817 and the ab initio prediction $R_{np}^{\rm th}(^{208}{\rm Pb}) = 0.14-0.20$\,fm\,\cite{hu21}. This theoretical calculation is ultimately based on a wide range of chiral nuclear forces, which are associated with soft equations of state. Despite the large central value of $R_{np}^{\rm exp}(^{208}{\rm Pb})$ (and the associated large value of the nuclear symmetry energy slope parameter $L$) from PREX-II, the sizeable uncertainty in the measurement only eliminates the softest equations of state from our modeling. The most likely probability region for the radius of a $1.4\,\msun$ neutron star in the case of the maximally stiff high-density extrapolation is somewhat larger than the case of the smooth extrapolation. This is due to the fact that many equations of state that are relatively soft for densities $n<2n_0$ can nevertheless survive the likelihood functions that favor stiff equations of state due to the large increase in pressure after the transition density $n_c$. One therefore finds that smaller radii for $\sim 1.4\,\msun$ neutron stars are allowed in the maximally-stiff high-density extrapolation. The upper bound on the radii of $\sim 1.4\,\msun$ neutron stars is almost identical $R_{1.4} < 12.8$\,km under the two high-density EOS scenarios.

\begin{figure*}[t!]
    \centering
    \includegraphics[width=0.49\textwidth]{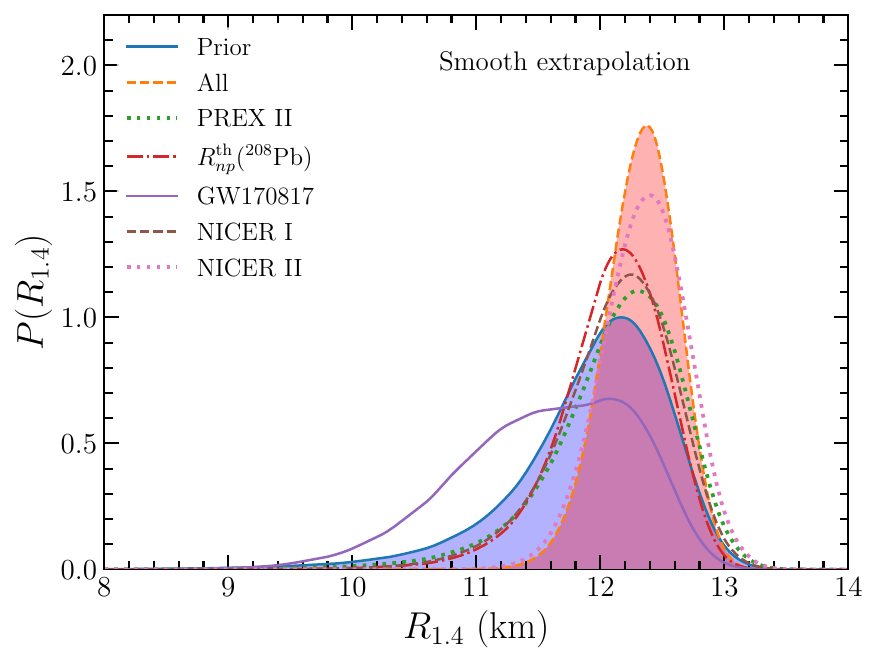}
    \includegraphics[width=0.49\textwidth]{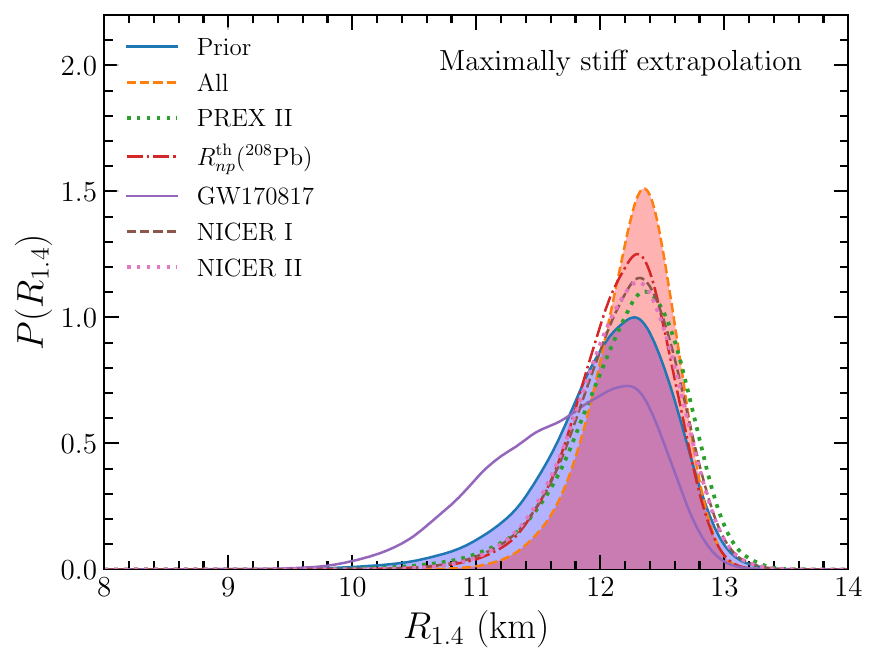}
    \caption{The $R_{1.4}$ probability densities (priors, individual posteriors, and combined posteriors) from the (a) smooth high-density extrapolation and (b) the maximally-stiff high-density extrapolation energy density functionals.}
    \label{fig:r14prob}
\end{figure*}
\begin{figure*}[t!]
    \centering
    \includegraphics[width=0.49\textwidth]{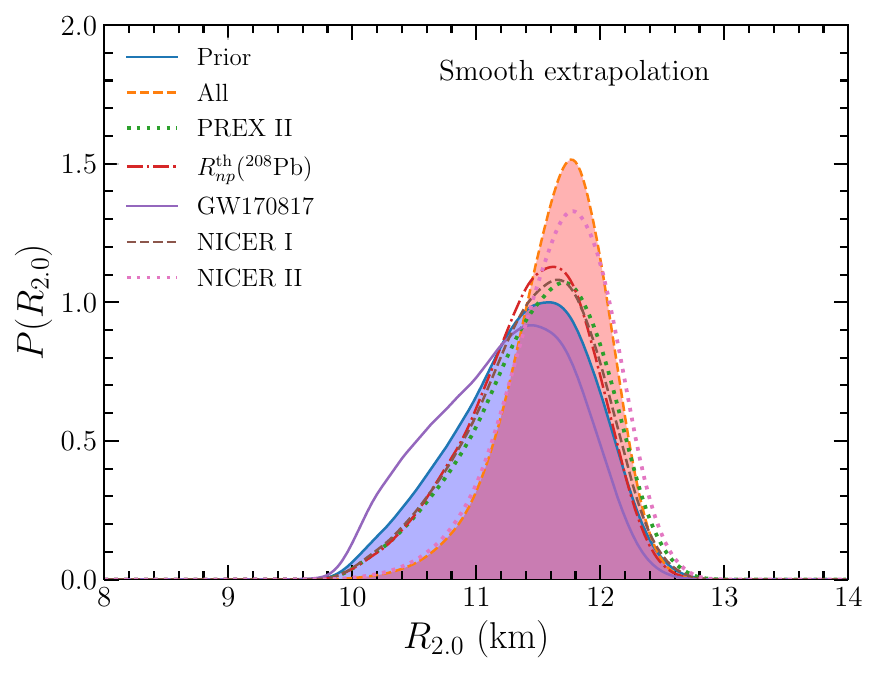}
    \includegraphics[width=0.49\textwidth]{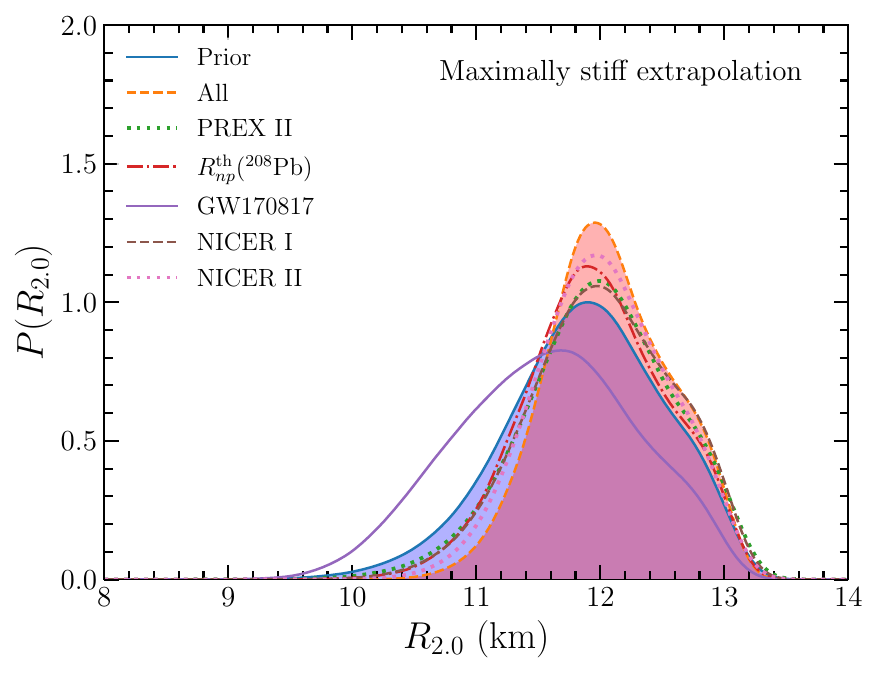}
    \caption{The $R_{2.0}$ probability densities (priors, individual posteriors, and combined posteriors) from (a) the smooth high-density extrapolation and (b) the maximally-stiff high-density extrapolation energy density functionals.}
    \label{fig:r20prob}
\end{figure*}

In Figures \ref{fig:r14prob} and \ref{fig:r20prob} we show the $R_{1.4}$ and $R_{2.0}$ probability densities when the smooth high-density extrapolation (left) and maximally-stiff high-density extrapolation (right) are employed. The blue shaded regions represents the prior distributions, while the orange shaded regions represents the final posterior distributions when all likelihood functions are combined. The other curves include only the effects of single likelihood functions (labeled in the figure) and thus can be used to evaluate the relative strengths of the individual experimental, theory, and observational constraints. From Figures \ref{fig:r14prob} and \ref{fig:r20prob} it is evident that GW170817 strongly favors soft equations of state, significantly shifting the distributions towards smaller radii for both $1.4\,\msun$ and $2.0\,\msun$ neutron stars. In the left panels of Figures \ref{fig:r14prob} and \ref{fig:r20prob}, one also sees the importance of heavy neutron stars and the NICER II constraint when our nuclear theory models are extrapolated without modification into the high-density regime. The NICER II observation has less impact in the maximally-stiff high-density scenario since equations of state that are soft around $n\sim 2n_0$ can be made sufficiently stiff at high densities to support heavy neutron stars. Instead, for the maximally-stiff high density assumption, the PREX-II neutron skin thickness constraint plays the largest role in eliminating soft equations of state. These observations highlight the general fact that the impact of heavy neutron star observations on theoretical modeling depends sensitively on the choice of the high-density equation of state.

\begin{table}[t]
\begin{center}
	\caption{The 68\% and 90\% credibility ranges for $R_{1.4}$ associated with the smooth and maximally-stiff high-density priors as well as the two posteriors accounting for all likelihood functions included in this work.}
	\begin{tabular}{ccc}
		\hline
                           & ~$R_{1.4}(68\%)$[km]~   & ~${R}_{1.4}(90\%)$\,[km]~  \\
		\hline
		Prior (Smooth)     &  $12.16^{+0.34}_{-0.78}$ & $12.16^{+0.59}_{-1.42}$   \\
		Prior (Stiff)      &  $12.28^{+0.26}_{-0.71}$ & $12.28^{+0.49}_{-1.2}$    \\
		Post.\ (Smooth)    &  $12.38^{+0.23}_{-0.35}$ & $12.38^{+0.39}_{-0.57}$     \\
		Post.\ (Stiff)     &  $12.36^{+0.22}_{-0.44}$ & $12.36^{+0.38}_{-0.73}$ \\
		\hline
\end{tabular}
\label{tb:r14}
\end{center}
\end{table}

\begin{table}[b]
\begin{center}
	\caption{The 68\% and 90\% credibility ranges for $R_{2.0}$ associated with the smooth and maximally-stiff high-density priors as well as the two posteriors accounting for all likelihood functions included in this work.}
	\begin{tabular}{ccc}
		\hline
                           & ~$R_{2.0}(68\%)$[km]~   & ~${R}_{2.0}(90\%)$\,[km]~  \\
		\hline
		Prior (Smooth)     &  $11.58^{+0.35}_{-0.79}$ & $11.58^{+0.61}_{-1.19}$   \\
		Prior (Stiff)      &  $11.90^{+0.64}_{-0.67}$ & $11.90^{+0.97}_{-1.15}$ \\
		Post.\ (Smooth)    &  $11.76^{+0.27}_{-0.50}$ & $11.76^{+0.46}_{-0.84}$ \\
		Post.\ (Stiff)     &  $11.96^{+0.65}_{-0.40}$ & $11.96^{+0.94}_{-0.71}$ \\
		\hline
\end{tabular}
\label{tb:r20}
\end{center}
\end{table}

In Table \ref{tb:r14} we show the most probable $R_{1.4}$ values, the 68\% credibility intervals, and the 90\% credibility intervals for the prior and posterior probability distributions assuming either the smooth or maximally-stiff high-density prior. When we employ the smooth high-density extrapolation, we see that the 90\% uncertainty spread is reduced from 2.01\,km (prior) to 0.96\,km (posterior), which represents a dramatic reduction in the allowed radii of $1.4\,\msun$ neutron stars. This is due to competing likelihood functions, some of which favor stiff and others that favor soft equations of state. However, the 90\% credibility range of $R_{1.4} = 12.38^{+0.39}_{-0.57}$\,km is sufficiently large that there is no need to introduce phase transitions or other exotic hypotheses to account for all present data. For the maximally-stiff high-density extrapolation, the 90\% credibility range for the prior spans $1.69$\,km while the posterior is reduced to $1.11$\,km. The combined set of likelihood functions result in a nearly equal range of allowed radii in the two posteriors, with the maximally-stiff extrapolation allowing for a somewhat larger set of soft equations of state around twice saturation density.

In Table\,\ref{tb:r20} we show the numerical values for the $R_{2.0}$ probability distribution. As expected, small radii are disfavored mostly due to the PSR J0740+6620 radius analysis in our NICER II likelihood function. The width for the $90\%$ credibility range decreases from 1.80\,km (2.12km) to 1.30km (1.65km) in case of the smooth (maximally stiff) high-density extrapolation. The uncertainty in the 90\% width for $R_{2.0}$ is larger compared to the $R_{1.4}$ uncertainty width due to uncertainties in the dense matter composition and equation of state at the largest baryon number densities.

\begin{figure*}[t]
    \centering
    \includegraphics[width=0.49\textwidth]{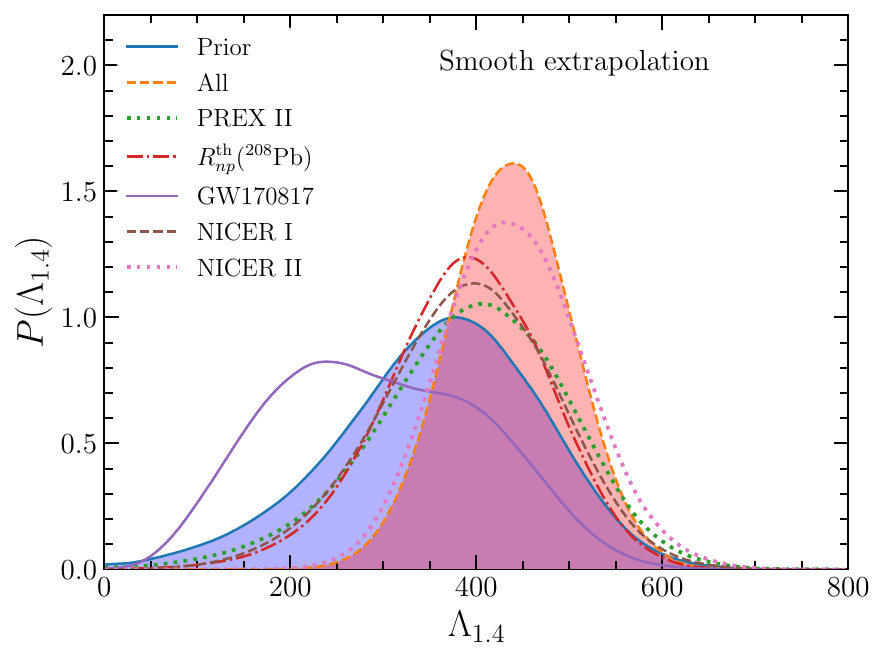}
    \includegraphics[width=0.49\textwidth]{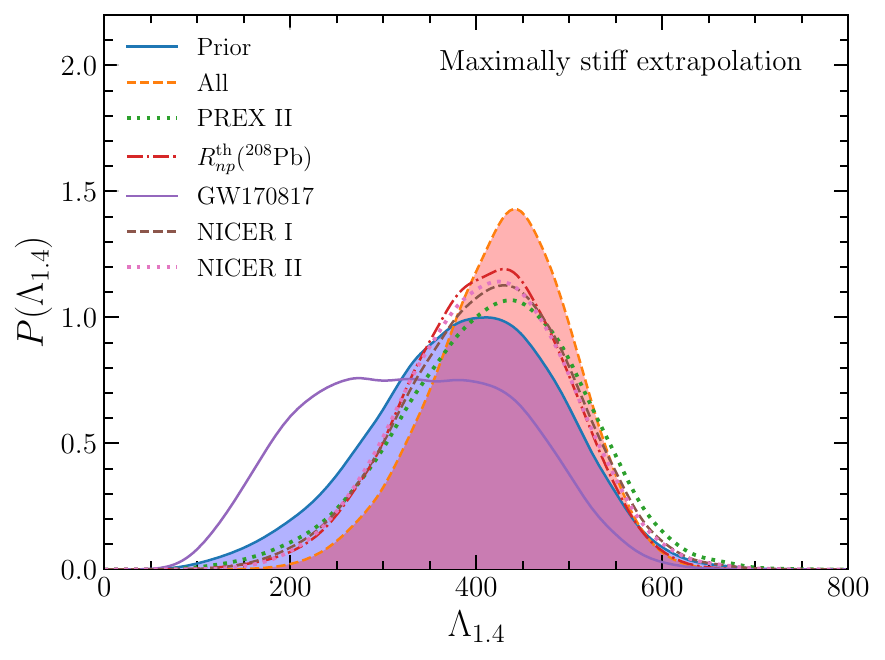}
    \caption{The $\Lambda_{1.4}$ probability densities (priors, individual posteriors, and combined posteriors) from (a) the smooth high-density extrapolation and (b) the maximally-stiff high-density extrapolation energy density functionals.}
    \label{fig:lambda14}
\end{figure*}

\begin{table}[b]
\begin{center}
	\caption{The 68\% and 90\% credibility ranges for $\Lambda_{1.4}$ associated with the smooth and maximally-stiff high-density priors as well as the two posteriors accounting for all likelihood functions included in this work.}
	\begin{tabular}{ccc}
		\hline
                           & ~$\Lambda_{1.4}(68\%)$~   & ~$\Lambda_{1.4}(90\%)$~  \\
		\hline
		Prior (Smooth)     &   $376^{+91}_{-129}$ & $376^{+151}_{-216}$   \\
		Prior (Stiff)      &   $408^{+81}_{-123}$ & $408^{+136}_{-195}$ \\
		Post.\ (Smooth)    &   $440^{+60}_{-71}$ & $440^{+101}_{-113}$  \\
		Post.\ (Stiff)     &   $440^{+61}_{-89}$ & $440^{+103}_{-144}$  \\
		\hline
\end{tabular}
\label{tb:l14}
\end{center}
\end{table}

Figure \ref{fig:lambda14} shows the prior density (blue shaded band), individual posterior densities (lines), and combined posterior density (red shaded band) for the tidal deformability of a $1.4\,\msun$ neutron star using both the smooth extrapolation prior (left) and the maximally stiff extrapolation prior (right). In general, the different likelihood functions have qualitatively similar effects on neutron star tidal deformabilities and radii (shown in Figure \ref{fig:r14prob}). This is due to the positive correlation between $\Lambda_{1.4}$ and $R_{1.4}$ as seen in Figure \ref{fig:corrori}. In particular, the likelihood from GW170817 strongly favors low values for $\Lambda_{1.4}$ in both prior scenarios since it is more difficult to deform neutron stars with smaller radii. The ab initio theory prediction for the neutron skin thickness $R_{np}^{\rm th}(^{208}{\rm Pb})$ gives only a small shift in the peak of the $\Lambda_{1.4}$ distribution, since it is strongly consistent with our chiral effective field theory constraints on the nuclear equation of state. The other constraints (NICER I, NICER II, PREX-II) all give larger shifts toward higher tidal deformabilities. Interestingly, the large uncertainty in the PREX-II likelihood function leads to greater statistical weight at both high and low values of $\Lambda_{1.4}$ compared to most of the other likelihood functions. Overall, the maximally-stiff high-density extrapolation exhibits a wider range in $\Lambda_{1.4}$ than the smooth extrapolation case, but the two posteriors peak at similar values of $\Lambda_{1.4}\sim 440$. We also note that most likelihood functions have a smaller impact on the $\Lambda_{1.4}$ distribution in the maximally-stiff high-density extrapolation compared to the smooth extrapolation prior because of the greater flexibility in the maximally-stiff model. Nevertheless, for both priors the combined effects of all likelihoods is to shift the $\Lambda_{1.4}$ distribution to larger values as seen by comparing the blue-shaded and orange-shaded curves.

\begin{figure*}[t!]
    \centering
    \includegraphics[width=0.49\textwidth]{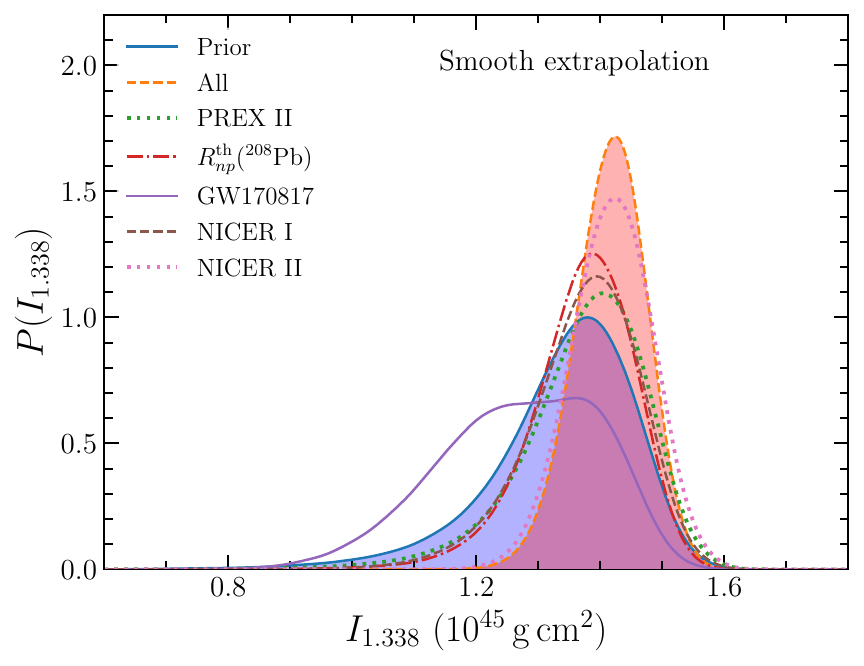}
    \includegraphics[width=0.49\textwidth]{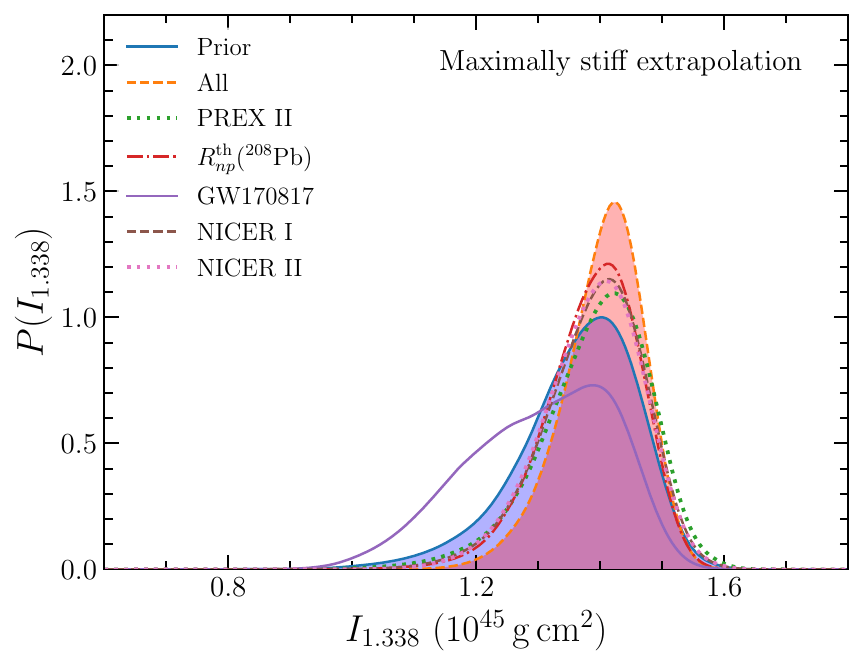}
    \caption{The $I_{1.338}$ probability densities (priors, individual posteriors, and combined posteriors) from (a) the smooth high-density extrapolation and (b) the maximally-stiff high-density extrapolation energy density functionals.}
    \label{fig:moi1338}
\end{figure*}

Table \ref{tb:l14} shows the most likely value of $\Lambda_{1.4}$ as well as the 68\% and 90\% credibility ranges under the smooth and maximally-stiff high-density extrapolations. The two prior probability distributions are already strongly consistent with the constraint $\Lambda_{1.4}=190^{+320}_{-120}$ from GW170817 \cite{abbott2018b}. The largest difference is that the analysis of GW170817 cannot strongly constrain the lower bound of the tidal deformability, whereas chiral effective field theory calculations exclude the extremely soft equations of state that would be needed to reach values of the tidal deformability $\Lambda_{1.4}<150$. The 90\% credibility ranges for the smooth and maximally-stiff high-density priors have uncertainties of 370 and 330, respectively, whereas the posteriors have uncertainties of 210 and 250. Again, we find that the smooth high-density extrapolation has a larger spread in the prior but a smaller spread in the posterior, primarily due to the strong constraints from NICER-II and the existence of massive $\sim 2\,\msun$ neutron stars. Interestingly, the two posteriors have the same peak value of $\Lambda = 440$ and quite similar uncertainties, indicating that the posterior is not dominated by the choice of prior.

In Figure \ref{fig:moi1338} we show the prior density (blue shaded band), individual posterior densities (lines), and combined posterior density (red shaded band) for the moment of inertia $I_{1.338}$ of PSR J0737-3039A using both the smooth high-density extrapolation (left) and the maximally stiff high-density extrapolation (right). The neutron star moment of inertia is strongly correlated with both the radius and tidal deformability, and therefore the trends observed in the two latter cases (see Figures \ref{fig:r14prob} and \ref{fig:lambda14}) are also exhibited by the moment of inertia. Like the radius $R_{1.4}$ and tidal deformability $\Lambda_{1.4}$, the probability distribution of $I_{1.338}$ is not symmetric about its peak value but instead has a longer tail toward small values. The ab initio theory prediction for the $^{208}$Pb neutron skin thickness eliminates a small fraction of the softest equations of state and stiffest equations of state such that the peak of the distribution is nearly the same as that of the prior. Comparatively, the PREX-II neutron skin thickness measurement gives rise to longer tails at both low and high values of the moment of inertia. Both posteriors, however, have very similar peak values and uncertainty widths.

\begin{table}[b]
\begin{center}
	\caption{The 68\% and 90\% credibility ranges for $I_{1.338}$ (in units of $10^{45}\,\rm{g\,cm}^{2}$) associated with the smooth and maximally-stiff high-density priors as well as the two posteriors accounting for all likelihood functions included in this work.}
	\begin{tabular}{ccc}
		\hline
                           & ~$I_{1.338}(68\%)$~   & ~$I_{1.338}(90\%)$~  \\
		\hline
		Prior (Smooth)     &   $1.375^{+0.068}_{-0.143}$ & $1.375^{+0.118}_{-0.260}$   \\
		Prior (Stiff)      &   $1.400^{+0.056}_{-0.132}$ & $1.400^{+0.100}_{-0.221}$\\
		Post.\ (Smooth)    &   $1.430^{+0.039}_{-0.071}$ & $1.430^{+0.074}_{-0.115}$ \\
		Post.\ (Stiff)     &   $1.425^{+0.041}_{-0.089}$ & $1.425^{+0.074}_{-0.146}$\\
		\hline
\end{tabular}
\label{tb:i1338}
\end{center}
\end{table}

In Table \ref{tb:i1338} we show the most likely value of $I_{1.338}$ as well as the 68\% and 90\% credibility ranges under the smooth and maximally-stiff high-density extrapolations. Recently, Kramer {\it et al.} \cite{kramer2021} have obtained the value $I_{1.338}<3\times 10^{45}\,\rm{g\,cm}^{2}$ at the 90\% credibility level for the moment of inertia of PSR J0737-3039A. We see that all of our models satisfy this upper bound on the moment of inertia and in fact lie much below the empirical upper bound. Landry and Kumar \cite{landry18} obtained $I_{1.338}=1.15^{+0.38}_{-0.24}\times 10^{45}{\rm g\,cm}^2$ from neutron star universal relations \cite{yagi13} linking the neutron star tidal deformability and moment of inertia. Recently, Grief {\it et al.} \cite{greif20} obtained $1.058<I_{1.338}<1.708\times 10^{45}{\rm g\,cm}^2$ from the speed of sound model and piecewise polytropic model constrained by chiral effective field theory and neutron star observables. Compared with all of these works, our posterior analysis gives a considerably smaller 90\% credibility width of $0.19$ and $0.22 \times 10^{45}{\rm g\,cm}^2$ from the smooth extrapolation and maximally-stiff extrapolation posteriors, respectively. The probability density for $I_{1.338}$ peaks around $1.43\times 10^{45}\,\rm{g\,cm}^2$ under both high-density models. The net effect of all likelihood functions is to reduce the uncertainty while only slightly increasing the peak of the $I_{1.338}$ distribution.

\section{Discussion}

\label{sec:diss}

We have studied the impact of recent theoretical and experimental investigations of the neutron skin thickness of $^{208}$Pb on predicted neutron star radii, tidal deformabilities, and moments of inertia. These have been included in a comprehensive Bayesian statistical analysis that also includes constraints from chiral effective field theory, properties of medium-mass and heavy nuclei, and astronomical observations of neutron star radii and tidal deformabilities. We have employed two choices for the high-density equation of state and find that the associated posteriors are very similar and do not depend sensitively on this choice. We find that the ab initio nuclear theory calculation of $R_{np}^{\rm th}(^{208}{\rm Pb})$ narrows the uncertainty in neutron star properties without strongly modifying the peak probabilities for radii, deformabilities, and moments of inertia. In comparison, the PREX-II experimental determination of $R_{np}^{\rm exp}(^{208}{\rm Pb})$ leads to broader posterior distributions for neutron star properties that are peaked at larger central values but also have longer tails toward the minima of the distributions. Given the large allowed space of posterior equations of state from the smooth high-density extrapolation scenario, there is no need to introduce exotic degrees of freedom or phase transitions in neutron stars to accommodate all current constraints. Our present modeling can be further refined through future measurements of neutron-rich nuclei at rare-isotope beam facilities, ab initio nuclear theory calculations, and neutron star observations.

\vspace{6pt} 



\authorcontributions{Conceptualization, Y.L and J.H.; methodology, Y.L and J.H.; software, Y.L.; validation, Y.L and J.H.; formal analysis, Y.L and J.H.; investigation, Y.L and J.H.; resources, Y.L and J.H.; data curation, Y.L.; writing---original draft preparation, Y.L.; writing---review and editing, Y.L and J.H.; visualization, Y.L and J.H.; supervision, Y.L and J.H.; project administration, Y.L and J.H.; funding acquisition, Y.L and J.H.. All authors have read and agreed to the published version of the manuscript.}

\funding{This research was funded by Ewha Womans University Research Grant of 2021(1-2021-0520-001-1), the National Research Foundation of Korea (NRF) grant funded by the Korea government (MSIT) (No. 2021R1A2C2094378), the US National Science Foundation Grant No.\ PHY1652199 and the U.\ S.\ Department of Energy National Nuclear Security Administration Grant No.\ DE-NA0003841.}

\dataavailability{Data from this study will be provided upon request by Y.L.}

\acknowledgments{We thank Baishan Hu and Christian Forss\'en for providing posterior equation of state samples associated with the theoretical determination of the $^{208}$Pb neutron skin thickness. Portions of this research were conducted with the advanced computing resources provided by Texas A\&M High Performance Research Computing.}

\conflictsofinterest{The authors declare no conflict of interest.} 




\appendixtitles{no} 

\begin{adjustwidth}{-\extralength}{0cm}

\reftitle{References}

\end{adjustwidth}
\end{document}